\documentclass[preprint,amsmath,amssymb,superscriptaddress]{revtex4}
\usepackage{graphicx}  
\usepackage{dcolumn}   
\usepackage{bm}        
\usepackage{bbm}       
\usepackage{amsmath, amssymb, graphics}
\usepackage[normalem]{ulem}   
\newcommand{\bra}[1]{\langle#1\vert} 
\newcommand{\ket}[1]{\vert#1\rangle} 
\newcommand{\beq}{\begin{equation}}
\newcommand{\eeq}{\end{equation}}
\newcommand{\bea}[1]{\begin{equation}\begin{array}{#1}}
\newcommand{\eea}{\end{array}\end{equation}}
\newcommand{\beqn}{\begin{eqnarray}}
\newcommand{\eeqn}{\end{eqnarray}}
\newcommand{\hq}[0]{\hat{q}}
\newcommand{\bq}[0]{\bar{q}}
\newcommand{\hH}[0]{\hat{H}}

\usepackage{color}
\definecolor{red}{rgb}{0.9,0.0,0.1}
\definecolor{green}{rgb}{0.2,0.8,0.1}

\begin{document}
\newcommand{\ave}[1]{\langle #1 \rangle}
\newcommand{\hx}[0]{\hat{x}}
\title{Monitoring the wave function by time continuous position measurement} 
\author{Thomas Konrad$^1$, Andreas Rothe$^1$, 
  Francesco Petruccione $^1$ and Lajos Di\'osi$^2$\\
$^1$ Quantum Research Group, School of Physics, University of KwaZulu-Natal, Private Bag
  X54001, 4000 Durban, South Africa
\\
$^2$Research Institute for Particle and Nuclear Physics, 
H-1525 Budapest 114, P.O.Box 49, Hungary}

\begin{abstract}
We consider a single copy of
a quantum particle  moving in a potential and show that it is possible
to monitor its  complete wave function by only continuously measuring
its position. While we assume that the potential is known, no
information is available about its state initially. In order to monitor
the wave function, an estimate of the wave function is propagated due to
the influence of the potential and continuously updated according to
the results of the position measurement. We demonstrate by numerical simulations that the
estimation reaches arbitrary values of accuracy below $100\%$ within a
finite time period for the potentials we study. In this way our method grants, a certain time after the
beginning of the measurement, an accurate real-time record of the state
evolution including the influence of the continuous
measurement. Moreover, it is robust against sudden perturbations of the system 
as for example random momentum kicks from environmental particles,
provided they occur not too frequently. 
\end{abstract}
\maketitle
Monitoring - continuous observation - of a
  dynamical system in the presence of randomness is employed not only in
  physics and chemistry, e.g., to survey the motion of comets, the growth of thin layers
  or the dynamics of chemical reactions. It is a sub-discipline of
  robotics and also plays a vital role
  in 
  other fields such as earth sciences and aeronautics
  with numerous applications from climate observation, control of
  robots and vehicles
to remote sensing. 
Monitoring tasks can be modeled by stochastic processes supported 
by continuous updates of estimates according to the observed random data
- also called stochastic filtering \cite{KalBuc61}. 
A special challenge, however, is posed
in the realm of quantum physics; the preparation - not to mention
monitoring or control - of
individual atoms, electrons and photons 
remained experimentally unattainable for half a century. The theory
of quantum monitoring only emerged 20 years ago \cite{Bel88,Dio88,WisMil93,Car93}. 
Nowadays, however, nano-technology and quantum information processing 
strongly inspire a mathematical theory of monitoring and control of single
quantum degrees of freedom like, e.g., the position of an atom or
a nano-object. A more challenging aim is  to monitor
  and control the entire state of individual quantum systems.   

Monitoring a quantum system encounters principle 
difficulties that lie in the
characteristic traits of quantum nature itself: incompatible observables
such as position and momentum and irreversible state change introduced
by measurements. Methods have been developed to employ
  monitoring in order to determine the pre-measurement state
  \cite{SilJesDeu05}, for parameter estimation \cite{ChaGer08}, to
  track Rabi oscillations \cite{AudKonScher01, AudKonScher02, AudKleeKon07} and
  -combined with feedback- for cooling purposes
  \cite{Stecketal06} or to reach a targeted state
  \cite{ShaJac08}. Moreover, the possibility of state
  monitoring has been studied for special systems \cite{DohTanParWal99,OxtGamWis08}. We are here going to show, that monitoring the position 
of a single quantum particle promises - via our theory -  
the monitoring of the full wave function, i.e., the complete
  state of a particle, and demonstrate 
its power by numerical simulations. 
Like always
in the quantum realm, monitoring will unavoidably alter the original 
(unmonitored) evolution of the wave function. Strong monitoring assures 
very robust fidelity of the estimated wave function but it has little
to do with the unmonitored wave function. Fortunately, in many cases, 
a suitable strength of monitoring assures both robust fidelity and 
slight change of self-dynamics. Needless to say, that such compromise is
not due to any weakness in our theory. It is definitely an ultimate 
necessity enforced by the Heisenberg uncertainty relations. 

In the following the concept of continuous observation is interpreted as 
the asymptotic limit of dense sequences of unsharp position measurements on a
single quantum particle. We describe the inference of its
wave function from the sequence of the measured position data
and eventually compare the true wave function with its estimate using
simulations of quantum particles moving in various potentials. Among
them is the H\'enon-Heiles potential, which in classical
physics implies chaotic behaviour and thus exposes tracking of
dynamics to extreme conditions.            
\section{Monitoring the position}
Time-continuous position measurement can be understood as an idealization
of a sequence of discrete unsharp position measurements  carried out 
consecutively on a single copy of a quantum particle \cite{Dio88}.
The notion of unsharp measurement is instrumental here.
Such an unsharp measurement of the position $\hq$ can be realized as 
indirect von Neumann measurement; instead of measuring the particle's position 
directly, an ancilla system is scattered off the particle and
then the ancilla is measured \cite{Neu55,CavMilb87, Busch95, BrePet02,JacSte06}. The observed results yield limited information on the position $\hq$ of the scatterer. 
In a simple description, a single unsharp measurement of resolution $\sigma$ collapses the 
wave function onto  a neighborhood with 
characteristic extent $\sigma$ of a random value $\bq$:
\begin{equation}
\psi(q)\longrightarrow 
\frac{1}{p(\bq)}\sqrt{G_\sigma(q-\bq)}\psi(q)\,,
\label{Gpsi}
\end{equation}
where $G_\sigma(q)=(1/\sqrt{2\pi\sigma^2})\exp(-q^2/2\sigma^2)$ is a central
Gaussian function.
The random quantity $\bq$ is the measured position which determines the
collapse, i.e., the weighted projection, of the wave function. The probability to obtain
  the measurement result $\bq$ - which also
plays the role of the normalization factor of the post-measurement wave function -
reads:
\begin{equation}
p(\bq)= 
\int G_\sigma(q-\bq)\vert\psi(q)\vert^2 dq\,.
\label{pbq}
\end{equation}
As a matter of fact, sharp (direct) von Neumann position measurements are 
the idealized special case while unsharp measurements - though not 
necessarily with the Gaussian profile - are the ones which we encounter in 
practice and which suite  a tractable theory of real-time monitoring of 
the position of a single quantum particle. 

In our discretized model of monitoring a single particle, we assume an
unknown initial wave function $\psi_0(q)$ and 
we are performing consecutive unsharp position measurements of 
resolution $\sigma$ at times $t=\tau,2\tau,\dots$, resp., yielding the 
corresponding sequence $\bq_t$ of measurement outcomes.
Between two consecutive unsharp measurements the wave function 
evolves according to its Schr\"odinger equation (self-dynamics).

The resolution $\sigma$ and the frequency $1/\tau$ of unsharp 
measurements should be chosen in such a way as not to heavily distort the 
self-dynamics of the particle. It turns out that the relevant parameter
is $\sigma^2\tau$, we call
\begin{equation}
\gamma=\frac{1}{\sigma^2\tau} 
\label{gamma}
\end{equation}
the strength of position monitoring. If $\sigma_\psi$ stands for the 
characteristic extension of the wave function, then 
$\gamma\sigma_\psi^2$ is the average decoherence rate at which
our monitoring distorts the monitored particle's self-dynamics.
We should keep this rate modest compared to the rate of the Schr\"odinger
evolution due to the Hamiltonian $\hH$ of the monitored
particle. Low values of the strength $\gamma$ may, however,
result in low efficiency of position monitoring and slow convergence
of our method  of wave function estimation, cp. Sec \ref{numSim}.  
The above constraints on $\sigma^2\tau$ can in general be matched
with further ones - see Sec.\ IV - that assure the applicability of
the continuous limit and its analytic equations. 

\section{Monitoring the wave function}
While it seems plausible that after a  sufficiently long  time $t$
the sequence of unsharp position measurements provides  enough data to 
estimate $\vert\psi_t(q)\vert^2$,
 it may come as  surprise that position measurements enable a  faithful
monitoring of the full wave function $\psi_t(q)$ as well. 
Let's just outline the reason. 
Measuring the   position  $\hq$ at  times 
$t=\tau,2\tau,\dots$ on an system with evolving Schr\"odinger wave function 
$\psi_t$  is equivalent to consecutive measurements of the Heisenberg 
observables $\hq_t=\exp(it\hH)\hq\exp(-it\hH)$ on a system with static wave function
$\psi_0$. The set of Heisenberg coordinates $\{\hq_t\}$ will exhaust
 a 
sufficiently large space of incompatible observables so that their 
measurements will  lead to a faithful determination of $\psi_0$ and - this way - 
to our faithful determination of $\psi_t$ for long enough times $t$.
In the degenerate case $\hH=0$, monitoring turns out to be trivial:
For long enough times, a large number $t/\tau$ of unsharp position
measurements of resolution $\sigma$ is equivalent with a single sharp 
measurement of resolution $\sigma/\sqrt{t/\tau}$, position monitoring
 thus yields just preparation of a static sharply localized wave function 
- an approximate `eigenstate' of $\hq$.

Our monitoring of the wave function means a real-time
  estimation of it, where the quality of monitoring depends on the
  fidelity of the estimation.
We start from a certain initial estimate $\psi_0^e$  and
  simulate its evolution according to the self-dynamics of the
  particle, which is assumed to be known, until time $t=\tau$. Immediately after
we have learned the first position $\bq_\tau$ from the first measurement on
the particle, we update the estimate according to the same
  rule (\ref{Gpsi}) as the actual wave function of the particle and
  renormalize it:
\begin{equation}
\psi^e_\tau(q)\longrightarrow 
\mbox{normalization}\times\sqrt{G_\sigma(q-\bq_\tau)}\psi^e_\tau(q)\,.
\label{Gpsie}
\end{equation}
This update resembles the Bayes principle of non-parametric statistical 
estimation.
We repeat this procedure  for $t=2\tau,3\tau\dots$ to expect that
the estimated  $\psi^e_t$ and the observed wave function $\psi_t$ will converge!
A rigorous proof of convergence is missing. In the continuous limit,
nonetheless,  it has been proved for the general case \cite{DioKonSchAud06} - not excluding
the lack of convergence in specific degenerate cases when the set of
 Heisenberg observables
$\{\hq_t\}$ remains too narrow to determine $\psi_0$. This is the
case for example for a  two-dimensional separable dynamics
in two coordinates $\hat x,\hat y$,  where only the coordinate $\hat x$
is monitored.
Rather than pursuing the rigorous theoretical conditions of convergence
cf. \cite{HanMab05}, we turned to  numerical tests  of 
continuous measurements that have definitely
confirmed our method. 

\section{Numerical Simulations}\label{numSim}
We simulated the evolution of a single hydrogen atom
subject to continuous measurements in
several potentials. However, the conclusions of our discussion are
not
restricted to hydrogen atoms; similar results can be expected for
atoms with higher masses in appropriately scaled potentials.   

The coupled evolutions of wave function, measurement readout and the estimated wave function were simulated numerically by
discretizing the corresponding stochastic differential equations
(cp. Sec.\ref{Methods}). For this purpose we employed a corresponding scheme of
Kloeden and Platen which is accurate up to second order
in the time step of the discretization  \cite{KLoPla92,BrePet02}.

\begin{figure}[ttt]
\centering
\includegraphics[width=16cm]{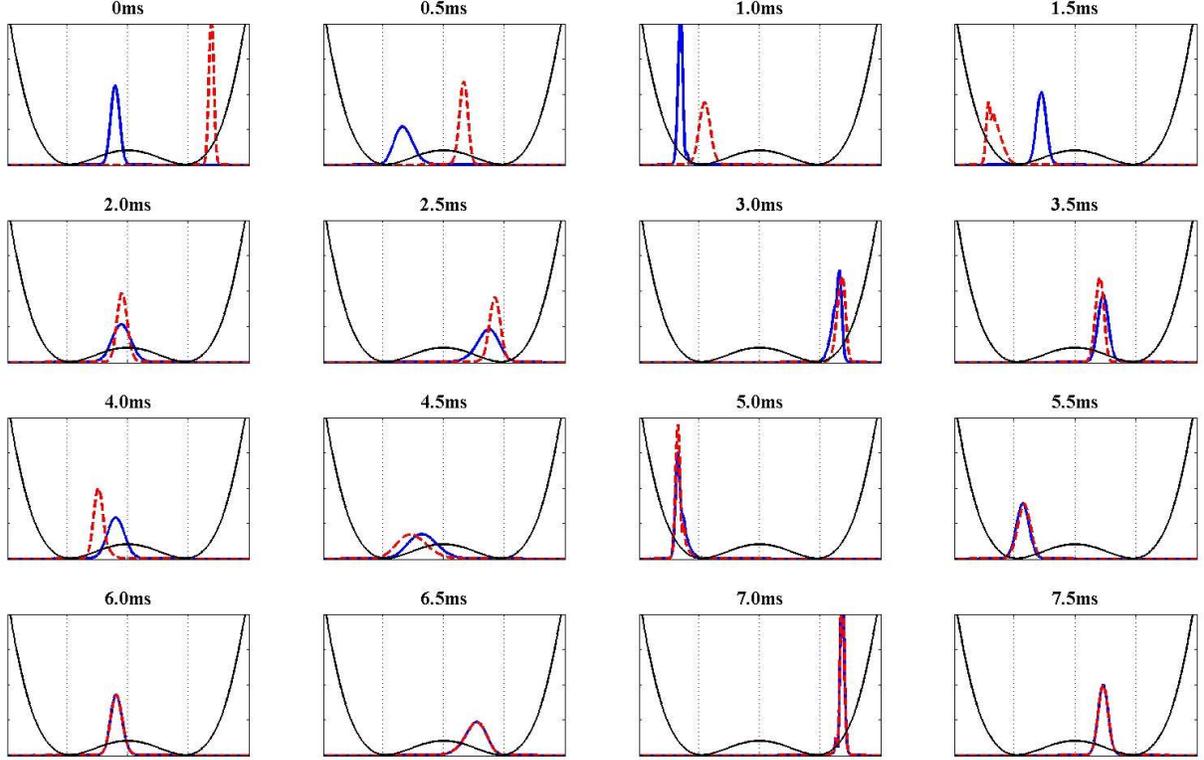}
\caption{Time sequence of real probability density of atomic position  $|\psi|^2$ (blue solid line) and the estimated probability
  density of position $|\psi^e|^2$ (read dashed line). The solid black
  line represents the double well potential as a function of
  position. Its minima are 189 $\mu$m apart and the height of the
  central maximum is given by $1\times 10^{-13}$ eV.}
\label{sequence}
\end{figure}

In order to study the relation between the evolution of the
wave function of the particle on one hand and the evolution of its 
estimate on the other hand we first restrict to a one-dimensional
spatial motion. In this case the graphical representation is the
simplest and thus gives a clear picture of the convergence between
real and estimated wave functions. As example we consider a hydrogen
atom situated in a quartic double well potential with a shape as depicted in
Fig.\ref{sequence}. 

We assumed a continuous measurement of the position of the
hydrogen atom with strength $\gamma=9.9856 /(\mu\mbox{m})^2\mbox{s}$. In
order to get an impression of the dimensions of the measurement, let us invoke Eq.~(\ref{gamma}) to 
note that this value of $\gamma$ may correspond, e.g.,  to single Gaussian
measurements with spatial resolution of $\sigma= 1.4$mm repeated at
time periods $\tau=50$ns. The spatial resolution of the single
weak measurements \cite{Dio06} is thus  $140$ times poorer than the
width   $\sigma=10 \mu$m of the 
initial  Gaussian wave function of the atom.       
In Fig.\ref{sequence} we have depicted snapshots recorded at
different times of the spatial probability density $|\psi(x)|^2$ of the
H-atom (blue solid line) and the squared modulus $|\psi^e(x)|^2$ of the estimated
wave function (red dashed line). 
 Initially both
probability densities assume the form of Gaussians which differ in
width and location within the double-well potential. The sequence of
pictures demonstrates the convergence of the densities in the course
of a continuous measurement. 

The real probability density $|\psi(x)|^2$  of the H-Atom, which
possesses initially a slightly higher mean energy than the middle peak of the potential,  oscillates back  and forth between the
sides  of the potential. The oscillatory motion of the centre of
$|\psi(x)|^2$  would also be expected qualitatively without measurements -as
well as from a classical particle of the same mass moving in the double
well. However, the real probability density does
not spread as it would do without measurements. This localisation
effect caused by the continuous unsharp position measurement adapts the motion of the H-Atom  to that of a classical particle which
is perfectly localised at each instance. This illustrates the
influence of the measurements and points to a particular kind of control of the
wave function that can be exercised by means of unsharp position
measurements. A smaller measurement strength would lead to less
disturbance of the unitary motion in the potential but also to less
speed of convergence between real and estimated probability density.       
 
The
estimated probability density  $|\psi^e(x)|^2$, which is centered
initially on the right-hand side of the potential, follows the real wave function until after
approximately one oscillation period the corresponding probability
densities coincide and evolve identically thereafter. But not only the
probabilities to find the atom at a certain position converge, in fact
the  complete wave function $\psi(x)$ and its estimate $\psi^e(x)$ coincide after a
sufficiently long period! It can be proved analytically, that the
estimation 
fidelity $F=|\langle \psi\ket{\psi^e}|$, which measures the overlap
between the wave functions $\psi$ and $\psi^e$, when averaged over many
realisations of the continuous measurement increases with time and
converges to $1$ \cite{DioKonSchAud06}. Numerical simulations for the
double-well potential show that in a typical realisation with initial
values as described above and measurement strength $\gamma=9.9856
/(\mu\mbox{m})^2\mbox{s}$  the fidelity amounts to more than $95\%$ after $1.5$  oscillation periods.

\begin{figure}[t]
\includegraphics[width=9.4cm]{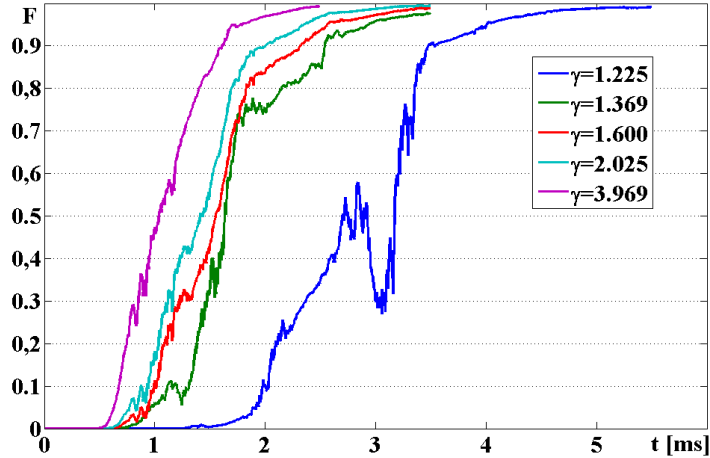}
\caption{The fidelity of estimation $F$ is plotted as a
  function of the duration of the continuous measurement of a H-Atom
  moving in a mexican hat potential for different
  values of measurement strength  $\gamma$ in units of $10/(\mu \mbox{m})^2\mbox{s}$. The height of the mexican hat's  central peak situated
  at the origin of the reference frame is given by $1.07\times 10^{-12}$eV,
  its minima lie on a concentric circle with radius $40\mu$m around
  the origin.  The wave function  $\psi(x)$ and its estimate
  $\psi^e(x)$ are initially Gaussians centered at ($-55\mu$m,$
  -14.8\mu$m) and ($-103.6\mu$m,$
  -103.6\mu$m) with widths of $10\mu$m as well as  $5\mu$m,
  respectively. The plots demonstrate that perfect fidelity is reached
  eventually for all considered measurements strengths. However, the
  convergence time decreases with increasing $\gamma$.}
\label{fidstrength}
\end{figure}

Fig.\ref{fidstrength} shows the evolution of estimation fidelity for
of a continuously measured H-atom moving in a plane under the influence
of a mexican-hat potential. The latter is a
rotationally symmetric version of the double well in two spatial
dimensions. For the sake of simplicity we assumed that both position coordinates are
simultaneously and independently measured with the same strength $\gamma$. Such a continuous
measurement of both coordinates typically yields evolutions of the estimation fidelities
which are shown in  Fig.\ref{fidstrength} for several values of the
measurement strength $\gamma$. In all depicted cases the fidelity comes
very close to one within a period of $5$ms, i.e., estimated and  real
wave function then coincide. Thereafter the 
dynamics of the wave function including the influence of the
measurement can thus be monitored with
perfect fidelity.

\begin{figure}[t]
\centering
\includegraphics[width=10cm]{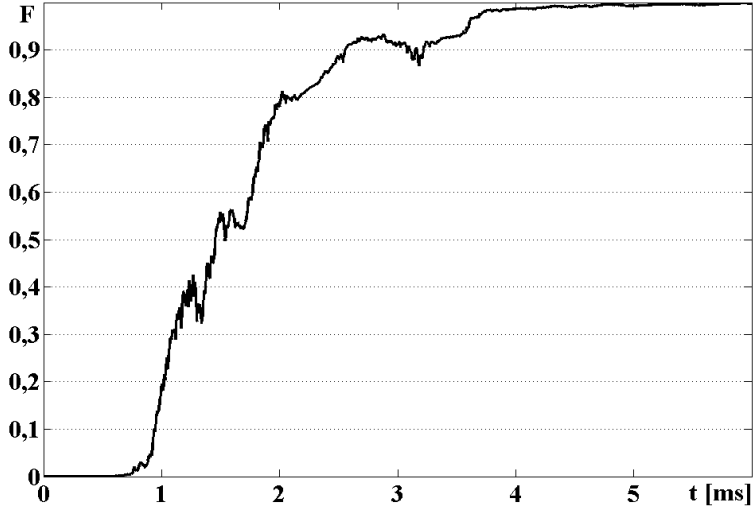}
\caption[Convergence for the H\'enon-Heiles potential]{The fidelity of estimation $F$ is plotted as a
function of the duration of the continuous measurement  at
  strength $\gamma=12.351/(\mu \mbox{m})^2\mbox{s}$
of a H-Atom in the non-integrable H\'enon-Heiles potential (\ref{Henon})
with parameters $A=5.44\times 10^{-17}\mbox{eV}/(\mu \mbox{m})^4,
a=13.09\mu \mbox{m}^2$ and $b=36.18\mu \mbox{m}$. 
 The wave function  $\psi(x)$ and its estimate
  $\psi^e(x)$ are initially Gaussians centered at ($-14.8\mu$m,$
  -29.6\mu$m) and ($-29.6\mu$m,$
  -29.6\mu$m) both with widths of $10\mu$m,
  respectively.
We find that the fidelity converges to 1 and therefore our estimate becomes a good approximation of the real wave function. 
}
\label{fig:henonconv}
\end{figure}

\begin{figure}[t]
\centering
\includegraphics[width=12cm]{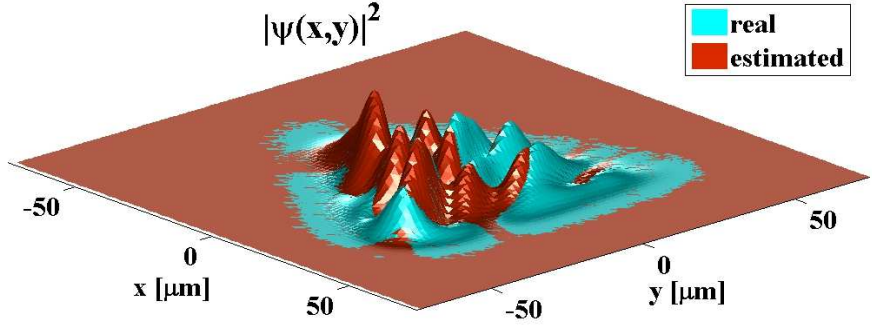}
\caption[HH2D]{The real (blue) and estimated (red) spatial probability densities $|\psi|^2$ and $|\psi^e|^2$, whichever is the bigger one, are depicted in the
H\'enon-Heiles potential after time $3.15$ms, at fidelity $0.9158$, for the same initial states as in Fig.\ref{fig:henonconv}.
} 
\label{fig:henon2Dplot}
\end{figure}

One might doubt that our monitoring remains efficient for heavily complex 
wave functions like those developing in classically chaotic systems. 
Instead of the integrable Mexican hat potential,
this time we  study the chaotic H\'enon-Heiles potential
which depends on the radius $r$ as well as on the azimuth $\phi$:
\begin{equation}
V(x,y)=A\left[r^4+ar^2+br^3\cos(3\phi)\right]\,. 
\label{Henon}
\end{equation}
For this potential we  simulated continuous position measurement and monitoring
  with the following results.
The saturation of fidelity is reassuring: the estimate converges 
to the real wave function (Fig.\ref{fig:henonconv}), which  is found at
 a  just slightly longer time scale than in the 
integrable Mexican hat potential, whereas the wave functions show an 
apparently irregular complex structure. 
 Fig.\ref{fig:henon2Dplot}  shows the estimate and the real wave functions
 with an average overlap (i.e. a fidelity) of  91.58\%. Indicating already an
 accurate overall estimation of the real wave function, this value still
 includes  in small areas of the potential differences of the  corresponding
 probability densities up to 35\% of their highest peak. In particular,
  Fig.\ref{fig:henon2Dplot} indicates  that faithful monitoring is
  not only possible when the shape of the wave function of the particle
  is close to a Gaussian but also for rather
  complex shapes.    

\begin{figure}[b!]
\centering
\includegraphics[width=14cm]{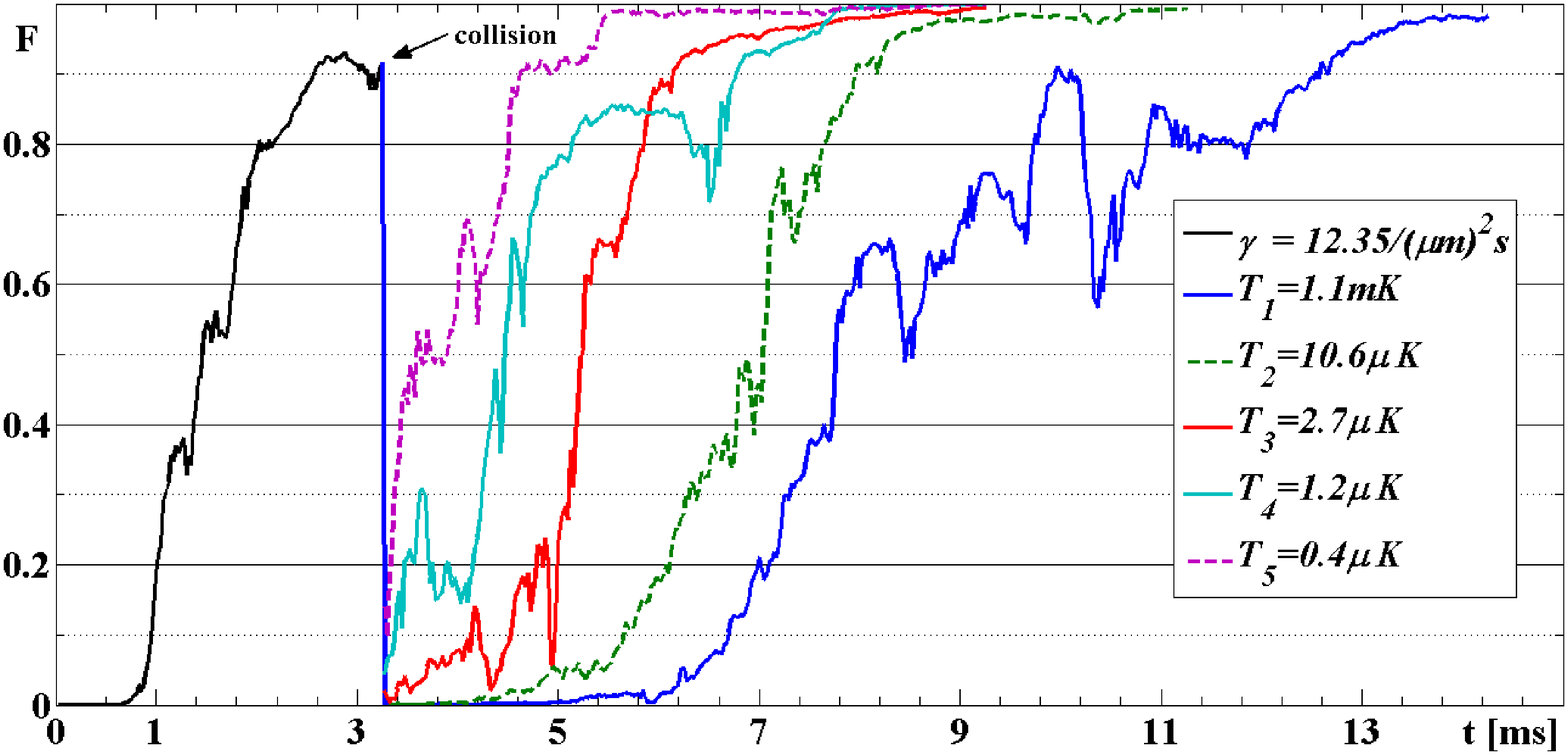}
\caption[Robustness of estimation scheme]
{At $t=3.15ms$ and fidelity $0.9158$, exactly when the snapshot of
  Fig.\ref{fig:henon2Dplot} was taken, 
the H-atom is hit by another thermal H-atom which causes an immediate drop of fidelity. We supposed a single momentum transfer $p_x=\sqrt{mk_BT}$
in the $x-$direction only, at different temperatures. The fidelity recovers and converges to $1$ after some time which depends on the
temperature, i.e., on the strength of the momentum kick.}
\label{fig:kick}
\end{figure}

Monitoring, i.e. continuous unsharp observation, has a specific 
capacity. It is its robustness against external unexpected perturbations. 
To demonstrate such a robustness, we assumed that close to saturation of the
estimation fidelity, like in Fig.\ref{fig:henon2Dplot}, our atom in the H\'enon-Heiles potential 
is suddenly perturbed, e.g., by a collision with an environmental 
  particle (here another hydrogen atom).
For simplicity, we assume a momentum kick $p_x$ along the x-direction
which 
implies a multiplication of the real wave function by the complex function $\exp(ip_x x/\hbar)$ 
hence the estimated wave function has to start a new cycle  of convergence. We map the 
momentum kick $p_x$ to a temperature by $k_BT=p_x^2/m$ as if it had a thermal
origin, just to give a hint of its strength. Numeric results (Fig.\ref{fig:kick})
show that the estimation fidelity recovers against these momentum
perturbations (cp.\ movie \cite{movie}). In realty,
repeated random perturbations might prevent perfect monitoring and fidelity will saturate
at less than $1$. This case is beyond the scope of our present work, its study
will be of immediate interest since real systems are subject to various noises
that are not measured at all. The monitoring theory  at non-optimum efficiency has been outlined earlier 
\cite{DioKonSchAud06}. 

\section{The Ito-method} \label{Methods}
The discrete sequence of unsharp measurements (\ref{Gpsi},\ref{pbq}) and wave function updates
(\ref{Gpsie}) possess their continuous limit \cite{Bar86} if we take $\tau\rightarrow0$ and 
$\sigma^2\rightarrow\infty$ at $1/\gamma=\tau\sigma^2=\mbox{const}$. 
In this `continuous limit' both the true wave function
$\psi_t(q)$ and the estimated wave function $\psi_t(q)$ become continuous
stochastic processes such that they are tractable by two stochastic 
differential equations respectively. The position measurement outcomes
$\bq_t$ do not yield a continuous stochastic process themselves. 
It is their time-integral $Q_t$, specified below, that becomes a continuous
stochastic process.

Let us consider the discrete increment of the true wave function during the period $\tau$,
cf. (\ref{Gpsi}). In Dirac formalism we get: 
\begin{equation}
\Delta\ket{\psi}=\exp(-i\tau\hH)\frac{1}{p(\bq)}\sqrt{G_\sigma(\hq-\bq)}\ket{\psi}-\ket{\psi}\,.
\label{Dpsi}
\end{equation}
For simplicity, we omit notations of time dependence $t$. The symbol
$\ave{\hq}$ stands for $\bra{\psi_t}\hq\ket{\psi_t}$.
In the continuous limit, the Eq.~(\ref{Dpsi}) transforms into the following Ito-stochastic differential equation \cite{Dio88}:
\begin{eqnarray}
d\ket{\psi}&=&\left(-i\hH-\frac{\gamma}{8}(\hq-\ave{\hq})^2\right)dt\ket{\psi}\nonumber\\
&{}&{} +\frac{\sqrt{\gamma}}{2}(\hq-\ave{\hq})(dQ-\ave{\hq}dt)\ket{\psi}\label{SSE}
\end{eqnarray}
The equation of the discrete increment $\Delta\ket{\psi^e}$ of the
estimate  (slight change) assumes the same form as Eq.~(\ref{Dpsi}) of $\Delta\ket{\psi}$ but
the normalization factor differs from $1/p(\bq)$, cf. (\ref{Gpsie}).  Yet, it yields the same Ito-stochastic 
differential equation as the equation above.
The estimated state $\ket{\psi_t^e}$ must be evolved according to the
same non-linear differential equation (\ref{SSE}) that describes the evolution of the monitored particle's state
$\ket{\psi_t}$.
These two equations are coupled via the stochastic process $Q$ whose discrete
increment is defined by $\Delta Q=q\tau$, in the continuous limit this means formally 
$Q_t=\int_0^t q_s ds$ where $q_s$ is the measured position at time $s$. In realty, the random process
$Q_t$ is obtained from the measured data $\{\bq_t\}$. If the measurement is just simulated, like in our work,
then in the continuous limit $\Delta Q$ transforms into the Ito-differential $dQ$ whose random evolution
can be generated by the standard Wiener process $W$ via $dQ=\ave{\hq}dt+\gamma^{-1/2}dW$.
Of course, $Q_t$ breaks the symmetry
between the stochastic processes $\psi_t$ and $\psi^e_t$ because $dQ/dt$ fluctuates around 
$\bra{\psi_t}\hq\ket{\psi_t}$ and not around $\bra{\psi^e_t}\hq\ket{\psi^e_t}$.  

The stochastic differential equation (\ref{SSE}) - combined with the same one for $\ket{\psi^e}$- is a suitable 
approximation of our discrete model (Secs. II-III) under two conditions: 
(i) a single measurement does not resolve any particular
structure of the wave function, i.e., $\sigma\gg\sigma_\psi$ where
$\sigma_{\psi}$ is the width of the spatial area on which $\psi$ is
not negligibly small. Thus $\sigma_\psi$ can, e.g., be of the order of
magnitude of the available width of the confining potential. 
(ii) The length of the time period $\tau$  between two
consecutive measurements is small compared to the timescale of self-dynamics
generated by the Hamiltonian $\hH$. 
Then the discrete model of position monitoring and wave function estimation 
becomes tractable by the time-continuous equation (\ref{SSE})
depending on the single parameter $1/\gamma=\tau\sigma^2$, cf. also Eq.~(\ref{gamma}).

\section{Summary}
We simulated numerically continuous position measurements carried
out on a single quantum particle in one- and two-dimensional potentials. In
order to monitor the evolution of an initially unknown state of the
particle in a known potential, we estimated its wave function and
updated the estimate continuously employing  the measurement
results. 

Our simulations show
that for all considered potentials the overlap between estimated and
real wave function comes close to $1$ after a finite period of
measurement -guaranteeing thereafter precise knowledge of the
particle's state and a real-time monitoring of its further
evolution with high fidelity. The power of our method is indicated by the
ability to monitor even the motion of a particle in a classically
chaotic potential  subject to continuous position measurement. 

We thus
demonstrated, that monitoring the complete state
of a quantum system with infinite dimensional state space is feasible
by continuously measuring a single observable 
on a single copy of the system. Moreover, the simulations indicate that our
monitoring method is robust against sudden external perturbations such
as occasional random momentum kicks. How much and what kinds of external noise
this monitoring scheme tolerates is important for its applicability in
control and error correction tasks, and might be object of future
research.         

\section{Acknowledgements}
We gratefully acknowledge support by the Bilateral Hungarian-South
African R\&D Collaboration Project, Hungarian OTKA grant 49384 and
South African NRF Focus Area grant 65579. We
thank J.\ Audretsch and A.\ Scherer for discussions. In particular, we
are grateful to Ronnie Kosloff for his idea to monitor chaotic
dynamics as well.

\bibliography{measurement}
\bibliographystyle{unsrt}

\end{document}